\documentclass[
reprint, letterpaper
superscriptaddress,
 amsmath,amssymb,
 aps,
pre,
]{revtex4-1}

\usepackage{microtype}
\usepackage[T1]{fontenc}

\usepackage{graphicx}
\usepackage{dcolumn}
\usepackage{bm}
\usepackage{amssymb}

\usepackage[dvipsnames]{xcolor}
\usepackage{hyperref}
\hypersetup{
	colorlinks=true,
	linkcolor=blue,
	filecolor=magenta,
	urlcolor=black,
	citecolor=blue,
}
\usepackage{braket}

\usepackage{float}

\begin{document}

\preprint{APS/123-QED}

\title{High-degeneracy Potts coarsening}

\author{J.\ Denholm}
\email{j.denholm@strath.ac.uk}

\affiliation{SUPA, Department of Physics, University of Strathclyde, Glasgow, G4 0NG, Scotland, UK}

\date{\today}

\begin{abstract}
I examine the fate of a kinetic Potts ferromagnet with a high ground-state degeneracy that undergoes a deep quench to zero-temperature. I consider single spin-flip dynamics on triangular lattices of linear dimension $8 \le L \le 128$ and set the number of spin states $q$ equal to the number of lattice sites $L \times L$. The ground state is the most abundant final state, and is reached with probability $\approx 0.71$. Three-hexagon states occur with probability $\approx 0.26$, and hexagonal tessellations with more than three clusters form with probabilities of $\mathcal{O}(10^{-3})$ or less. Spanning stripe states---where the domain walls run along one of the three lattice directions---appear with probability $\approx 0.03$. ``Blinker'' configurations, which contain perpetually flippable spins, also emerge, but with a probability that is vanishingly small with the system size.
\end{abstract}

\maketitle

\section{Introduction}
When a kinetic ferromagnet with non-conserved magnetisation undergoes a deep quench to zero-temperature, the final states are intriguingly diverse. After the ensuing coarsening regime, surviving domain structures compete for dominion over the final state. Na\"{i}vley, one could assume that a single domain will ultimately prevail as the ground state is necessarily reached---but this is far from the truth even in simple models.

In the nearest-neighbour Ising model of linear dimension $L$, the ground state is \emph{always} reached in one-dimension~\cite{krapivsky2010}, yet \emph{never} reached in three-dimensions~\cite{Olejarz_a_2011, Olejarz_b_2011}. In two-dimensions, the situation is markedly richer: only $62 \%$ of realisations proceed directly to the ground state on a timescale of $\mathcal{O}(L^{2})$~\cite{Spirin_a_2001, Spirin_b_2001}. Surprisingly, $34 \%$ of trajectories become trapped in frozen two-stripe states, which are infinitely long-lived and form on a timescale of $\mathcal{O}(L^{2}\ln L)$~\cite{Spirin_a_2001, Denholm_2020}. The remaining $4\%$ of instances reach ephemeral diagonal winding configurations, which ultimately collapse to homogeneity on a timescale of $\mathcal{O}{\left(L^{3.5}\right)}$~\cite{Spirin_a_2001, Spirin_b_2001}.

The explanation underpinning these final states is an apparent one-to-one mapping with the equivalent crossing probabilities of critical continuum percolation~\cite{Arenzon_2007, Barros_2009, Olejarz_2012}. Once a percolating domain structure has formed, the fate of the zero-temperature Ising model is sealed~\cite{Arenzon_2007, Barros_2009, Olejarz_2012}. Percolation and the domain growth in bi-dimensional coarsening have been readily studied~\cite{Arenzon_2007, Barros_2009, Olejarz_2012, Blanchard_2013, Cugliandolo_2016, Blanchard_2017_a, Corberi_2017, Tartaglia_2018, Insalata_2018, Bray1991, Bray1993, Sicilia_2007}.

After understanding the fate of the zero-temperature Ising model, it is natural to consider a system of greater degeneracy and to therefore study the dynamics of the $q$-state Potts model. Interest in the kinetics of Potts ferromagnets has been motivated by their utility in understanding coarsening in soap froths~\cite{Glazier_1990, Thomas_2006, Weaire_2009}, magnetic grains~\cite{Srolovitz_1984, Fradkov_1994, Raabe_2000, Zollner_2011}, natural tilings~\cite{Mombach_1990, Mombach_1993, Hocevar2010}, superconductors~\cite{Prozorov_2008} and magnetic domains~\cite{Babcock_1990, Jagla_2004}.

The domain growth in coarsening Potts systems has been extensively studied and is well understood~\cite{Safran_1982, Safran_1983, Sahni_1983_a, Sahni_1983_b, Grest_1988, Ferrero_2007, Oliveira_2009, Srolovitz_1985, Holm_1991, Petri_2008, Loureiro_2010, Loureiro_2012, Corberi_2019}. The existence of non-ground final states after a zero-temperature quench was realised in the Potts model before the equivalent finding in the Ising model~\cite{Oliveira_2004_a, Oliveira_2004_b, Olejarz_a_2013}. Nevertheless, there is an apparent lack of literature examining the late-time configurations that persist after a zero-temperature quench. In fact, there are seemingly only two studies that explicitly focus on this problem\nobreakdash---both of which are in two dimensions~\cite{Olejarz_a_2013, Denholm_2019}.

\begin{figure}[b!]
    \includegraphics[width=\columnwidth]{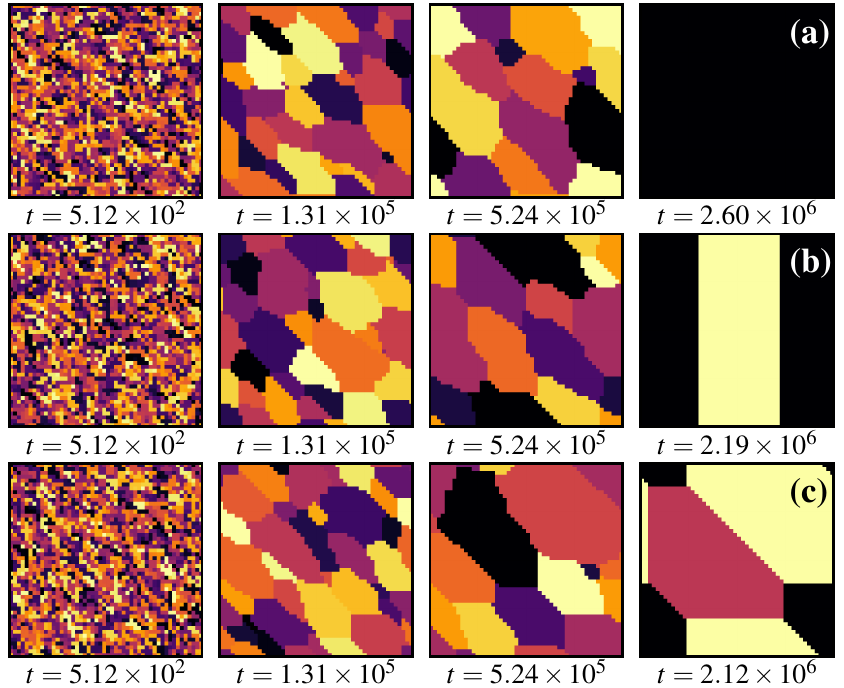}
    \caption{Snapshots of zero-temperature coarsening in a triangular-lattice Potts ferromagnet of $q=L\times L=2500$ for realizations reaching (a) ground, (b) two-stripe and (c) three-hexagon states. Distinct clusters are labelled by colour.}
    \label{fig:snapshots}
\end{figure}
On the \emph{square} lattice with $q = 3$, several oddities emerge: the ground state probability is only $\approx 0.1$~\cite{Olejarz_a_2013}, and the most prevalent final states are ``frozen'' configurations with two or more surviving spin states~\cite{Olejarz_a_2013}.  Surprisingly, ``blinker'' configurations also emerge, where the system forever wanders at constant energy~\cite{Olejarz_a_2013}. The strangest feature however is that of ``pseudo-blinkers'': after exorbitant time periods strongly resembling blinkers, single energy-lowering flips trigger sudden ``energy avalanches'', where the system suddenly declines in energy and macroscopically reorders~\cite{Olejarz_a_2013}. Identifying these configurations was a non-trivial numerical challenge~\cite{Olejarz_a_2013}.

On the \emph{triangular} lattice with $q=3$, the ground state is reached approximately $3/4$ of the time, and both three-hexagon and two-stripe states appear (FIG.~\ref{fig:snapshots}~(a)--(c))~\cite{Denholm_2019}. Blinkers seem only to occur only in small system sizes with $q>4$, and play a negligible role in the dynamics. The disparity between the fates of the square and triangular lattice Potts models at zero-temperature is surprising considering the affinity in the equivalent Ising models.

The origin of the contrasting fates of the square- and triangular-lattice Potts models is rooted in the behaviour of so-called ``T-junctions''. T-junctions are formed by the meeting of three domain structures comprised of different spin-types~\cite{Olejarz_a_2013}. At the centre of the junction, one finds spins which are trapped in local energy minima that cannot flip. The \emph{square}-lattice, by virtue of geometry, imposes the restriction that T-junctions are fixed in their spatial location. Thus, the only way for the system to escape is through some macroscopic disruption of the configuration---which is not always possible.~\cite{Olejarz_a_2013}.

However, on the \emph{triangular} lattice, the geometrical constraint on the location of T-junctions is lessened, and the centre of the junctions can move. Consequently, the system can escape from these configurations~\cite{Denholm_2019}. Perhaps the simplicity that emerges on the triangular lattice makes it a better candidate for one-day achieving the exact computation of the final states probabilities of the three-state Potts model at zero-temperature?

 Further study of the triangular lattice Potts model is interesting for a number of reasons. In the Ising model, we see two categories of final state, both of which span the linear dimension of the system. When we move to the three-state Potts model, a new topologically distinct final state emerges: the three-hexagon state. If we increase the degeneracy to $q=5$, blinker configurations, which are another fundamentally different final state, appear. One can therefore ask: are there other interesting features of the triangular lattice Potts model that yet remain uncovered? The general behaviour of the final state probabilities at high degeneracies is unknown.

 In this manuscript I examine the final state of a zero-temperature Potts ferromagnet with a high ground-state degeneracy on the \emph{triangular} lattice. I explore the special case where the number of spin states $q$ is equal to the number of lattice sites $L\times L$, which provides a natural upper bound on the number of spin states with considering. I detail the model and simulation method in Sec.~\ref{sec:methods}. In Sec.~\ref{sec:final-states}, I introduce the final states that emerge and estimate the frequency with which they occur. In Section~\ref{sec:time-evo}, I examine the number of clusters and surviving spins states as functions of time, before summarising my findings in Section.~\ref{sec:discussion}.

\section{Zero-temperature Potts model}
\label{sec:methods}

\begin{figure}[h!]
	\centering
	\includegraphics[width=0.75\columnwidth]{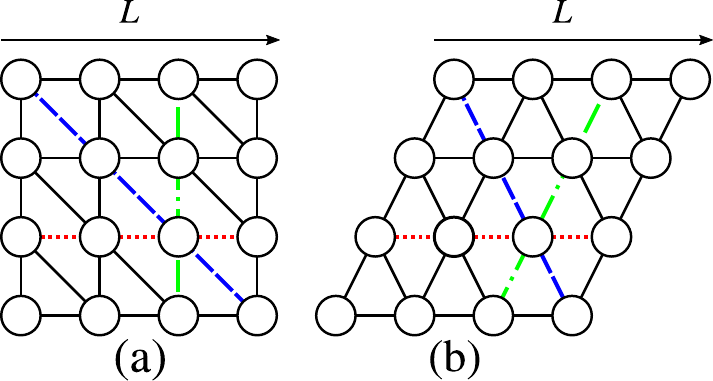}
	\caption{Equivalent triangular lattice geometries (a) and (b).}
	\label{fig:lattice_topology}
\end{figure}

I consider nearest-neighbour interactions on the triangular lattice geometry with periodic boundary conditions. I build the triangular lattice by taking a square lattice of length $L$ and adding diagonal bonds to the North-West and South-East (see FIG.~\ref{fig:lattice_topology}).

I initialise the system by placing each spin in a unique state, giving $q=L\times L$ states in total. The spin states are denoted by the integers $S_{i} \in \{1,2,\dots q\}.$ Like spins are said to be aligned and unlike spins misaligned. The total energy of the system is given by the Hamiltonian
\begin{equation}
\mathcal{H} = -2J\sum_{i, j}\big[\delta(S_{i},S_{j}) - 1\big]\,,
\end{equation}
where $J > 0$ is a ferromagnetic coupling constant, $\delta(S_{i}, S_{j})$ is the Kronecker delta and $j$ indexes the nearest-neighbours of each $S_{i}$. Thus, each misaligned neighbour provides an energy contribution of $+2J$.

To implement the dynamics, I employ continuous-time rejection-free kinetic Monte Carlo---where each spin is allowed to flip once, on average, in a single Monte Carlo time unit~\cite{Bortz_1975, Landau_2009, Sahni_1983_a, Hassold_1993}. This method is equivalent to the standard discrete time Monte Carlo method, where one allows $N=L\times L$ randomly selected spins the chance to flip once in a single time step~\cite{Bortz_1975, Landau_2009,Hassold_1993}. I endow the Hamiltonian with zero-temperature metropolis dynamics: energy lowering and energy conserving moves are accepted with probability $1$, while energy raising moves are forbidden~\cite{Sahni_1983_a, Hassold_1993}. The choice of dynamics is relatively flexible so long as one adheres to the principle of detailled balance, therefore one might also use Glauber dynamics.~\cite{Hassold_1993}.

The total rate, $r_{i}$, of spin $S_{i}$ is simply the sum of the transition probabilities over each of the $(q - 1)$ orientations it may flip to. Since I use zero-temperature Metropolis dynamics\nobreakdash---where the transition probabilities are $1$ or $0$\nobreakdash---the rate $r_{i}$ is simply a count of the number of transitions permitted by the dynamics. Let the total rate in the system be $R=\sum r_{i}$. To flip a spin, I select a site with probability $r_{i} / R$, draw randomly from its list of $r_{i}$ permissible transitions, and then flip the spin. Time advances as $\Delta t = -\log(u)\times (q - 1)/R$, where $u \in (0, 1)$ is a uniform random number and $\braket{-\log(u)} = 1$. I include a brief note on the simulation time in Appendix.~\ref{appendix:note-sim-time}.

\begin{figure*}[t!]
    \includegraphics[width=\textwidth]{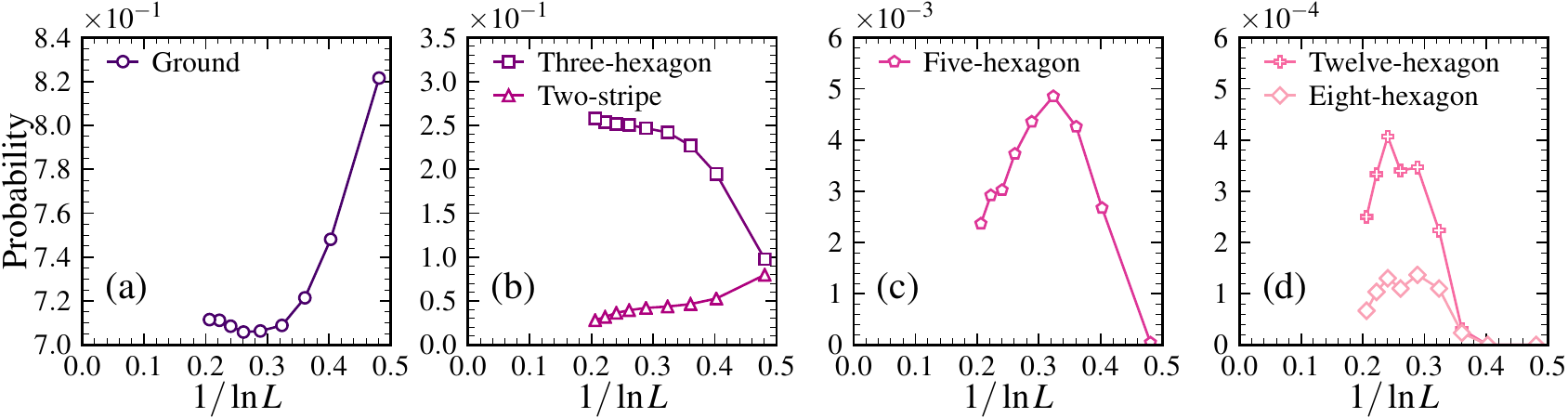}
    \caption{Probability of reaching: (a) the ground state; (b) a three-hexagon and two-stripe state; (c) a five-hexagon state; (d) a twelve- and eight-hexagon state. The data are based on $3\times10^{5}$ realisations.}
    \label{fig:probabilities}
\end{figure*}

An important simplifying feature at zero-temperature is that spins with no aligned neighbours may flip to $(q-1)$ other spin states, whereas spins with at least one aligned neighbour may only flip to align with other neighbouring spins. This consideration is useful when computing $r_{i}$, particularly at large $q$.

\section{Final states}
\label{sec:final-states}

I begin my examination of the final state probabilities with the ground state case, which I plot in FIG.~\ref{fig:probabilities}~(a). The ground state is the most abundant final state, and when $L=128$, it is reached with a probability of $\approx 0.71$. The ground state probability varies non-monotonically with $L$, making it difficult to obtain an asymptotic estimate. The non-monotonicity in the ground state probability is not unique to this system: it is also present in both the Ising model and small $q$ Potts models on the square and triangular lattice geometries~\cite{Spirin_a_2001,Spirin_b_2001, Olejarz_a_2013, Denholm_2019}.

\begin{figure}[t!]
    \centering
    \includegraphics[width=\columnwidth]{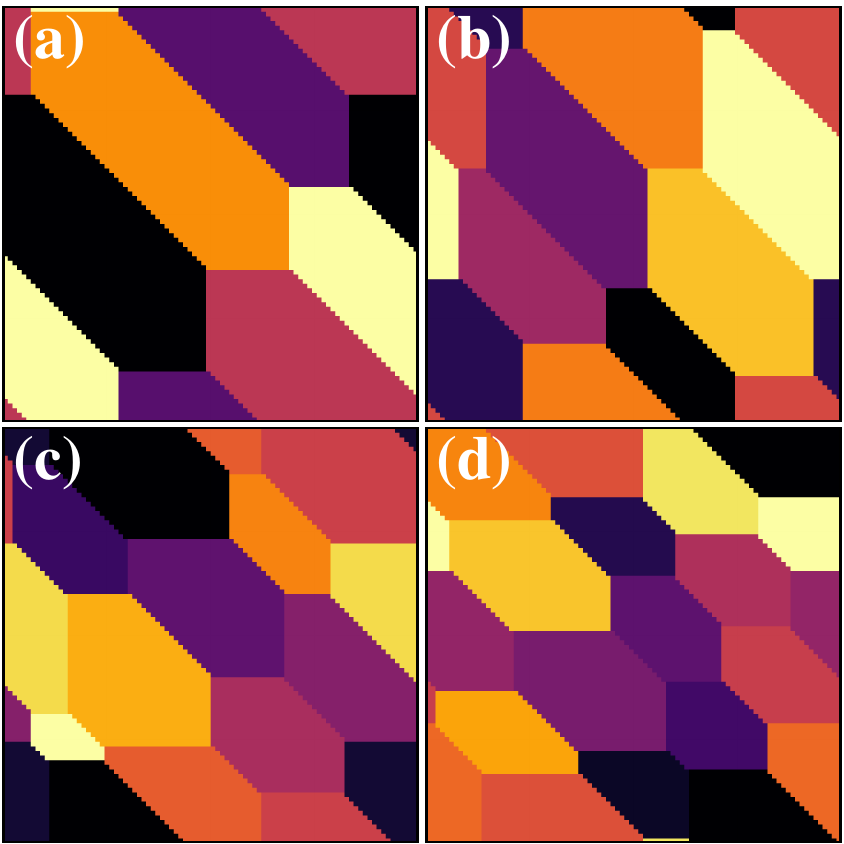}
    \caption{Frozen (a) five-, (b) eight-, (c) twelve- and (d) sixteen-hexagon states. Each cluster is labelled by colour and the lattice size is $L = 90$.}
    \label{fig:higher_hexagons}
\end{figure}

The next most common final states are three-hexagon and two-stripe states, which are reached with probabilities of $\approx 0.26$ and $\approx 0.03$ respectively (FIG.~\ref{fig:probabilities}~(b)). As well as three-hexagon states, I also find rarer subspecies of hexagonal tilings containing five, eight and twelve clusters\nobreakdash---examples of which are shown in FIG.~\ref{fig:higher_hexagons}~(a)--(c).

Hexagonal configurations with more than three clusters are much rarer than the three-cluster case; I plot the probability of finding five-, eight- and twelve-hexagon states in FIG.~\ref{fig:probabilities}~(c)--(d), showing they are orders of magnitude less abundant than their three-cluster counterparts. I also found a \emph{single} realisation which reached a \emph{sixteen}-hexagon state (see FIG.~\ref{fig:higher_hexagons}~(d)), making it the rarest of its kind.

The energy of an $n$-hexagon state depends only on the number of clusters, and not the individual cluster arrangement. Consider the three hexagon state in FIG.~\ref{fig:snapshots}~(c). The total length of interface between the domains is $3L$. There are spins on either side of these interfaces, giving $6L$ boundary spins. Each interface spin has two misaligned neighbours giving an energy contribution of $+4J$. Consequently, the total energy of any three-hexagon state is $24L$. I extend this reasoning to hexagon states with more than three clusters to obtain the energies shown in Table~\ref{tab:hex-energies}.

\begin{table}[b!]
    \begin{ruledtabular}
        \begin{tabular}{cccc}
            $n$   & Interface length  &  Interface spins  & Energy\\\hline
            $3$   & $3L$  &  $6L$     & $24L$\\
            $5$   & $4L$  &  $8L$     & $32L$\\
            $8$   & $5L$  &  $10L$    & $40L$\\
            $12$  & $6L$  &  $12L$    & $48L$\\
            $16$  & $7L$  &  $14L$    & $56L$
        \end{tabular}
    \end{ruledtabular}
    \caption{Energies of $n$-hexagon final states. Note: each bond is counted twice.}
    \label{tab:hex-energies}
\end{table}

\begin{figure}[t!]
    \centering
    \includegraphics[width=0.5\columnwidth]{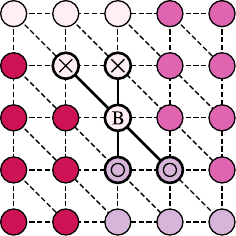}
    \caption{Zoom-in on a blinker spin (B) which can align with its North and North-west neighbours ($\times$) or its South and South-east neighbours ($\bigcirc$), but \emph{never} its East or West neighbours.}
    \label{fig:blinker-zoom}
\end{figure}

The scarcest final states are blinkers, which are configurations that contain perpetually active sites. ``Blinking'' spins flip eternally as they only have energy conserving transitions available to them. Consider the zoom-in on a blinker configuration in FIG.~\ref{fig:blinker-zoom}. When the spin is aligned with its North and North-west neighbours, it can only flip to align with its South and South-east neighbours, and vice versa. It can \emph{never} align with its East or West neighbours without raising its energy, which is forbidden at zero-temperature.

In small $q$ triangular-lattice Potts models, blinkers seemingly only occur when $q > 4$ with a probability that is vanishingly small with increasing $L$~\cite{Denholm_2019}. Here I find three main categories of blinkers: five-, eight- and twelve-cluster configurations. Each contain $\mathcal{O}(1)$ active sites which are pinned in the same way as the blinker spin in FIG.~\ref{fig:blinker-zoom}. The probability of reaching blinkers with five, eight and twelve clusters is $\mathcal{O}(10^{-3})$, $\mathcal{O}(10^{-4})$ and $\mathcal{O}(10^{-4})$ respectively. I also found a single realisation that reached a blinker state with sixteen clusters with $L=22$. The probability of reaching a blinker configuration---of any kind---on the triangular lattice is vanishingly small with increasing $L$.

\section{Time Evolution}
\label{sec:time-evo}
Two natural observables to consider when a high-$q$ Potts system is quenched are the number of clusters, $N_{c}$, and the number of extant spins states, $E_{q}$. A cluster is simply a group of aligned spins that are connected through nearest-neighbour contact, and the number of extant $q$ is a count of the distinct spin states that remain present in the system.

\begin{figure}[t!]
    \includegraphics[width=\columnwidth]{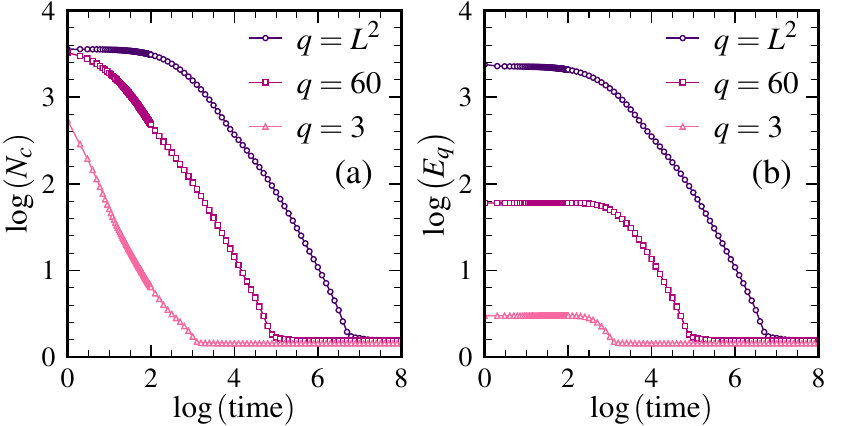}
    \caption{Time dependence of (a) the number of clusters $N_{c}$ and (b) the number of extant spin states $E_{q}$. The data are based on $10^{3}$ realisations with $L=60$.}
    \label{fig:time-evo}
\end{figure}

I compare the time evolution of these quantities for Potts systems with $q=3$, $q=60$ and $q=L^{2}$ in FIG.~\ref{fig:time-evo}.  As the number of spin states increases, the departure from the unmagnetised initial state slows---both $N_{c}$ and $E_{q}$ are increasingly stagnant at early times with greater $q$.

The explanation for this slow initial evolution is simple: consider the unmagnetised initial condition where no spin has any aligned neighbours. Each spin may freely undergo one of $(q - 1)$ possible transitions. In a small $q$ Potts system, the probability that a spin should flip to align with a neighbour is relatively high. However, when $q$ is large, the probability that a single flip should result in two neighbours aligning is small. Consequently, at large $q$, it takes longer for the spins to align, so the departure from the unmagnetised initial condition becomes increasingly slow.

\section{Discussion}
\label{sec:discussion}
I investigated the final state of a zero-temperature Potts ferromagnet where the ground-state degeneracy was equal to the number of lattice sites. With respect to $q=3$, the geometric and topological nature of the final states is not materially different, but their relative abundance \emph{is} different.

The ground state is the most prevalent final state, and is reached with probability $\approx 0.71$ when $L=128$. Three-hexagon states appear more frequently with increasing $L$, and are reached with probability $\approx 0.26$ when $L=128$. I also found hexagon states with five, eight, and twelve clusters with probabilities of $\mathcal{O}(10^{-3})$, $\mathcal{O}(10^{-4})$ and $\mathcal{O}(10^{-4})$ respectively. On-axis stripe states, where the domain walls run along one of the three lattices axes, also appeared. The probability of reaching two-stripe states decays with increasing $L$, and was $\approx 0.03$ when $L=128$. Blinker configurations with five, eight and twelve clusters also emerged. The probability of finding blinker configurations is $\mathcal{O}(10^{-3})$ or less, and is vanishingly small with increasing $L$.

The time evolution at high $q$ is inherently slow, meaning my probability estimates as $L\to\infty$ are necessarily crude. I illustrated this slow evolution by comparing the number of clusters and extant spin states as functions of time in Potts models with $q=3$, $q=60$ and $q=L^{2}=3600$.

There are a number of open questions concerning the fate of kinetic Potts ferromagnets at zero-temperature. The exact computation of the final state probabilities with $q=3$ has not yet been achieved. The connection with two-colour percolation that emerged in the Ising model enabled the precise conjecture of the final state probabilities\nobreakdash---perhaps a similar connection exists between the less-well-understood three-colour percolation and the three-state Potts model? Nevertheless, the affinity between the final states of the square- and triangular-lattice Ising models at zero temperature is not present in the equivalent Potts models, so the universality of a connection to three-colour percolation is unclear.

Furthermore, even if three-colour percolation does apply to a Potts ferromagnet with $q=3$, the question of $q>3$ remains; the general dependence of the final state probabilities on the number of spin states is still unknown. It is concevable that the high-$q$ limit will one day play a role in an analytical solution for the final state probabilities of the zero-temperature Potts model on the infinite triangular lattice geometry. In such a case, knowledge of how the final state probabilities behave in finite geometries, and of what kinds of final state to expect, will be important. The fact the final state probabilities for $q=L^{2}$ are different to the $q=3$ case is an interesting finding, and suggests the ground-state degeneracy plays an important role in determining the final state.

Additionally, the fate of the zero-temperature Potts ferromagnet on the simple cubic lattice remains unexplored. On the triangular lattice, where the spins have six nearest neighbours, the final states become materially simpler and more tractable. Perhaps the three-state Potts model on the simple cubic lattice, which also has a coordination number of six, will exhibit a similar simplicity?

\vspace{-\baselineskip}
\section*{Acknowledgements}
I thank Sid Redner for an interesting suggestion that sparked this work and Leticia Cugliandolo for encouraging comments. I acknowledge the ARCHIE-WeSt High Performance Computer based at the University of Strathclyde and grant EP/P015719/1 for computer resources. I also acknowledge EPSRC DTA5 grant EP/N509760/1 for financial support.

\bibliography{references.bib}

\appendix
\section{Note on simulation time}
\label{appendix:note-sim-time}
\begin{figure}[h!]
    \centering
    \includegraphics[width=\columnwidth]{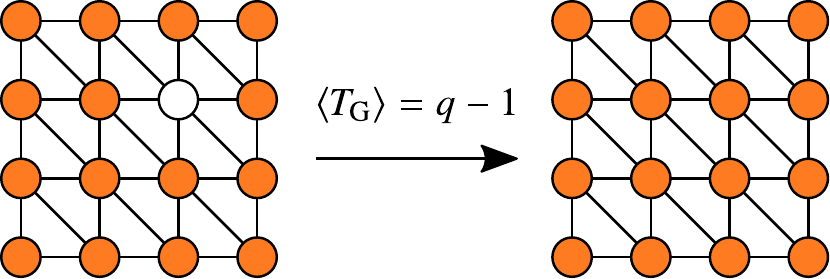}
    \caption{A realisation reaching the ground state at time $\braket{T_{G}}$.}
    \label{fig:time_cartoon}
\end{figure}

Consider a zero-temperature Potts system with only a single active site (see FIG~\ref{fig:time_cartoon}). If each site is allowed to flip once, on average, in a single time step, and sites with no aligned neighbours have $q - 1$ possible transitions available to them, the configuration in FIG.~\ref{fig:time_cartoon} reaches the ground state in $(q - 1)$ Monte Carlo time steps. When $q=2$, the ground state is reached in a single Monte Carlo time step, and when $q=3$, the ground state is reached in two Monte Carlo time steps, and so on. This feature significantly encumbers my simulations: say $L = 100$ and $q=L^{2}$; the active spin in FIG.~\ref{fig:time_cartoon} will flip $\mathcal{O}(10^{4})$ times before aligning with its neighbours.

\newpage~

\end{document}